\documentclass{emulateapj} 
\begin{document}

\newcommand{\cl}{{CL~0152.7-1357}}
\newcommand{\lynx}{{RDCS~0848+4453}}
\newcommand{\ms}{{MS~1054.4-0321}}
\newcommand{\V}{$V_{606}$}
\newcommand{\ra}{$r_{625}$}
\newcommand{\ia}{$i_{775}$}
\newcommand{\z}{$z_{850}$}

\newcommand{\etal}{{\em et~al.\,}} \title{The Possible $z=0.83$ Precursors of
  $z=0$, $M^{\star}$ Early-Type Cluster Galaxies\altaffilmark{1}} 
\altaffiltext{1}{Based on observations with the
NASA/ESA Hubble Space Telescope, obtained at the Space Telescope
Science Institute, which is operated by the Association of
Universities for Research in Astronomy, Inc., under NASA contract
No. NAS5-26555.  These observations are associated with programs
9919 and 9290.}

\author{B. P. Holden\altaffilmark{2}}

\author{M. Franx\altaffilmark{3}}

\author{G. D. Illingworth\altaffilmark{2}}

\author{M. Postman\altaffilmark{4}}

\author{J. P. Blakeslee\altaffilmark{5}}

\author{N. Homeier\altaffilmark{6}}

\author{R. Demarco\altaffilmark{6}}

\author{H. C. Ford\altaffilmark{6}}

\author{P. Rosati\altaffilmark{7}}

\author{D. D. Kelson\altaffilmark{8}}

\author{K.-V. H. Tran\altaffilmark{3,9}}

\altaffiltext{2}{UCO/Lick Observatory, University of California, Santa
  Cruz, 1156 High Street, Santa Cruz, CA 95065; holden@ucolick.org;
  gdi@ucolick.org} 
\altaffiltext{3}{Sterrewacht Leiden , Postbus 9513,
  NL= 2300 RA, Leiden,  Netherlands; franx@strw.leidenuniv.nl;
  vy@strw.leidenuniv.nl} 
\altaffiltext{4}{Space Telescope Science Institute, Baltimore, MD
  21218; postman@stsci.edu}
\altaffiltext{5}{Department of Physics and Astronomy, P.O. Box 642814,
  Washington State   University; Pullman, WA 99164; jblakes@wsu.edu}
\altaffiltext{6}{Department of Physics and Astronomy, Johns Hopkins
  University, Baltimore, MD 21218; nhomeier@pha.jhu.edu;
  demarco@pha.jhu.edu; ford@pha.jhu.edu} 
\altaffiltext{7}{European
  Southern Observatory, Karl-Schwarzschild-Strasse 2, D-85748 Garching
  Germany; prosati@eso.org} 
\altaffiltext{8}{Observatories of the
  Carnegie Institution of Washington, 813 Santa Barbara Street,
  Pasadena, CA 91101;   kelson@ociw.edu} 
\altaffiltext{9}{Harvard-Smithsonian Center for
  Astrophysics, 60 Garden Street, Cambridge MA 02138}

\shorttitle{The Possible Precursors of $z=0$ Cluster Early-Types}

\begin{abstract}

  We examine the distribution of stellar masses of galaxies in \ms\
  and \cl, two X-ray-selected clusters of galaxies at $z=0.83$.  Our
  stellar mass estimates, from spectral energy distribution fitting,
  reproduce the dynamical masses as measured from velocity dispersions
  and half-light radii with a scatter of 0.2 dex in the mass for
  early-type galaxies.  When we restrict our sample of members to high
  stellar masses, those over$> 10^{11.1}\ M_{\sun}$ ($M^{\star}$ in the
  Schechter mass function for cluster galaxies), we find that the
  fraction of early-type galaxies is 79\% $\pm 6$\% at $z=0.83$ and
  87\% $\pm 6$\% at $z=0.023$ for the Coma Cluster, consistent with no
  evolution.  Previous work with luminosity-selected samples has found
  that the early-type fraction in rich clusters declines from
  $\simeq$80\% at $z=0$ to $\simeq$60\% at $z=0.8$.  The observed
  evolution in the early-type fraction from luminosity-selected
  samples must predominantly occur among sub-$M^{\star}$ galaxies.  As
  $M^{\star}$ for field and group galaxies, especially late-types, is
  below $M^{\star}$ for cluster galaxies, infall could explain most of
  the recent growth in the early-type fraction.  Future surveys could
  determine the morphological distributions of lower mass systems
  which would confirm or refute this explanation.

\end{abstract}
\keywords{galaxies: clusters: general --- galaxies: elliptical and
lenticular, cD, --- galaxies: evolution --- galaxies: fundamental
parameters --- galaxies: photometry --- galaxies: clusters: \ms\ ---
\cl }

\section{Introduction}

The early-type galaxy fraction and morphology-density relation show
that the mix of galaxies we observe in clusters today has
significantly changed over the past 7-10 Gyrs
\citep{dressler97,lubin98,vandokkum2000,vandokkum2001,holden2004,smith2005,postman2005}.
This observation raises the question of what the descendants of
the additional late-type galaxies seen at higher redshifts are.
One possible approach to this problem is to compare the distribution
of masses of cluster galaxies at high and low redshifts. In this Letter
we use fundamental plane measurements and precise photometry to
estimate the mass-to-light ratios ($M/L$), and masses, of $z=0.83$
cluster galaxies and galaxies in the Coma Cluster.  When combined with
{\em Hubble Space Telescope (HST)} Advanced Camera for Surveys (ACS)
imaging, we have a unique data set for examining the masses and
morphologies of high redshift clusters galaxies, which we can compare
with a $z=0$ sample.  Throughout, we assume $\Omega_m =
0.3$, $\Omega_{\Lambda} = 0.7$ and $H_0 = 70\ {\rm km\ s^{-1}\
Mpc^{-1}}$.

\section{Data}
\label{data}

We use imaging from the {\em HST}/ACS to determine rest-frame optical
colors and morphological types.  Both \ms\ and \cl\ were observed with
F775W (hereafter \ia) and F850LP (hereafter \z) filters.  In addition,
\ms\ was observed with the F606W (hereafter \V) while \cl\ was imaged
in the F625W, or \ra, filter.  For a thorough discussion of the ACS
photometry, see \citet{blakeslee2005}.  \citet{postman2005} determined
the morphological types of the cluster members using the \ia\
data. \citet{postman2005} estimated the scatter on the early-type
fraction to be 6\% by comparing the fractions derived from different
classifiers.

We have 143 spectroscopically confirmed members in \ms\ with \ia $<
23.5$ mag \citep{tran2003}, with additional redshifts gathered since
that publication. For the second cluster in our sample, \cl\ at
$z=0.834$, we have 95 spectroscopic members with \ia $<23.5$ mag from
\citet{demarco2005a}.  We compute the completeness for each cluster
empirically as a function of \ia\ magnitude and morphological type as
was done by \citet[see Appendix B]{postman2005}.  We use these
completeness values as weights when computing derived quantities
below.

We compare the above data with the $U$, $B$, and $r$ photometry for
morphologically identified, spectroscopically confirmed Coma galaxies
from \citet{beijersbergen2002a,beijersbergen2002b}. Those authors
constructed a new sample of redshifts that is complete to $r <
16.27$ mag or $M_r < -18.75$ mag, and morphologically typed a large
number of previously unidentified galaxies.  We compute the rest-frame
$g-r$ color from the relation $g-r = 0.61 (B-r) - 0.07$, using
templates from \citet{bc03} and \citet{kinney96}, where $B-r$ comes
from the catalog of \citet{beijersbergen2002a}.

\section{Mass estimation}
\label{mass_est}

In a number of recent papers, galaxy masses have been estimated by
using observed spectral energy distributions.  The correlation
between rest-frame optical color and the mass-to-light ratio ($M/L$)
of the population for a variety of spectral types has been explored by
\citet{kelson2000c} and \citet{bell2001}.  Interestingly,
\citet{bell2003} and \citet{holden2006} found that fitting spectral
energy distributions to photometric data, in general yielded results
roughly equivalent to estimating  $M/L$ with only rest-frame
optical colors.

We use theoretical \citep{bc03} and empirical templates
\citep{kinney96} to interpolate between the observed ACS colors into
the rest-frame optical $g-r$ colors.  For \ms, we compute the
rest-frame as $g-r = 1.01 (i_{775}-z_{850}) - 0.05$ and $r = z_{850} -
0.32 (V_{606}-z_{850}) + 0.84$, while for \cl, we use $g-r = 1.00
(i_{775}-z_{850}) - 0.06$ and $r = z_{850} - 0.28 (r_{625}-z_{850}) +
0.63$.  \citet{bell2003} give two relations between the rest-frame
$g-r$ color and $M/L$ in the $r$ band, one from stellar population
models and one from the fundamental plane (FP) results of
\citet{bernardi2003}.  We use the relation derived from stellar
population models, but the results do not change significantly if we
use the relation from \citet{bernardi2003}.  We can also convert the
ACS photometry into the rest-frame $B-V$. We find that using the
rest-frame $B-V$ conversion to $M/L$ does not significantly change
our results.

\begin{figure}[htbp]
\begin{center}
\includegraphics[width=3.4in]{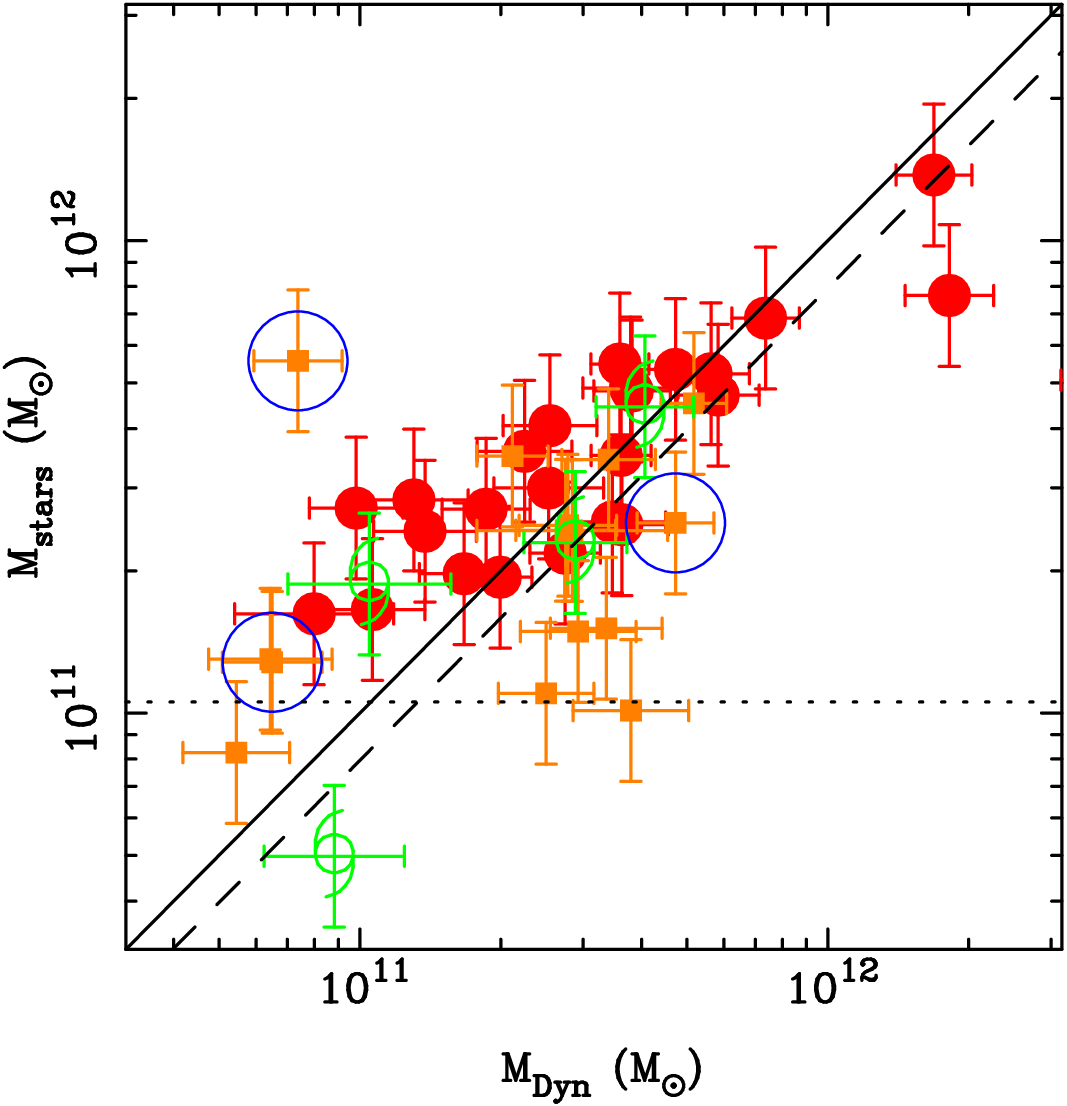}
\end{center}
\caption[fp_masses.ps]{ Mass derived from rest-frame optical
  colors compared with the dynamical mass for galaxies in \ms\ and
  \cl.  The symbols show elliptical galaxies as red circles, S0--S0/a's
  as orange squares, and Sa--Sbc as green spirals \citep[contains the
  classification]{postman2005}.  Those galaxies from \citet{wuyts2004}
  and \citet{postman2005} that are classified as in close pairs, or
  mergers, are enclosed by blue circles.  The dotted line shows the
  effective stellar mass limit assuming the colors of early-type
  galaxies in the clusters and the magnitude limit for the velocity
  dispersion samples.  In general, we find that the photometrically
  measured masses match the dynamically measured ones but with an
  offset of, on average, 0.13 dex.  The mean relation before the
  offset is applied we show with the dashed line, while the solid line
  shows a one-to-one agreement.  A smaller offset is required to match
  the Coma photometry to the Coma dynamical masses (see text for
  details).  We apply these offsets to both samples to ensure that the
  color-based mass estimates are consistent at high and low redshift.
}
\label{fp_masses}
\end{figure}

We have collected a sample of 51 galaxies with half-light radii and
velocity dispersions, 22 dispersions from \ms\ \citep{wuyts2004} and
29 dispersions in \cl\ \citep{jorgensen2005}, with sizes for both
clusters from \citet{holden2005b}.  In Figure \ref{fp_masses}, we
compare the masses derived from the relation $M = 5\sigma^2 r_e
/G$ or $M = 2*\log10 \sigma + \log10 r_e + 6.07 $ \citep{jfk96}, with
the masses estimated using the rest-frame $M/L_r$ values from ACS
colors, combined with our total magnitude estimates.  We find an
offset of 0.13 dex between our stellar mass estimates and the
dynamically estimated mass, shown in Figure \ref{fp_masses}.  The
scatter in the mass estimates for the elliptical galaxies is 60\%,
compared with 66\% for the whole sample.  It appears that, at lower
dynamical masses, our stellar masses are overestimates.  This is
likely a result of the luminosity limits used in selecting galaxies
for the velocity dispersion samples, which we illustrate with a dotted
line in Figure \ref{fp_masses}.  Future studies of the fundamental
plane to fainter magnitudes will test our color scaling of $M/L$
to fainter luminosities.

For the Coma data, we use the same conversion between $g-r$ and $M/L$
as for our two $z=0.83$ clusters.  Using the FP results of
\citet{jfk95sb} and \citet{jfk95vel}, we find good agreement between
our photometrically-estimated masses for the Coma cluster data and the
dynamical values, again with scatter of 46\% and 0.07 dex offset.  We
apply this offset to the Coma data.  Therefore, an object with the
same rest-frame $g-r$ and total $r$ magnitude will have the same mass
in both the high redshift and Coma samples.

We fit a Schechter function \citep{schechter76} to the distribution of
masses in Coma and find $M^{\star}$, the characteristic mass, to be
$10^{11.1 \pm 0.2}\ M_{\sun}$.  This value is higher than $10^{11.00}\
M_{\sun}$ (for $H_0 = 70\ {\rm km\ s^{-1}\ Mpc^{-1}}$) found in the
field survey of \citet{bell2003} because $L^{\star}$, which is $M_r =
-21.6$ mag AB for cluster galaxies
\citep{beijersbergen2002a,hansen2005} is brighter than $L^{\star}$,
$M_r = -21.3$ mag AB, for field galaxies found in \citet{bell2003}.

\begin{figure}[htbp]
\begin{center}
\includegraphics[width=3.4in]{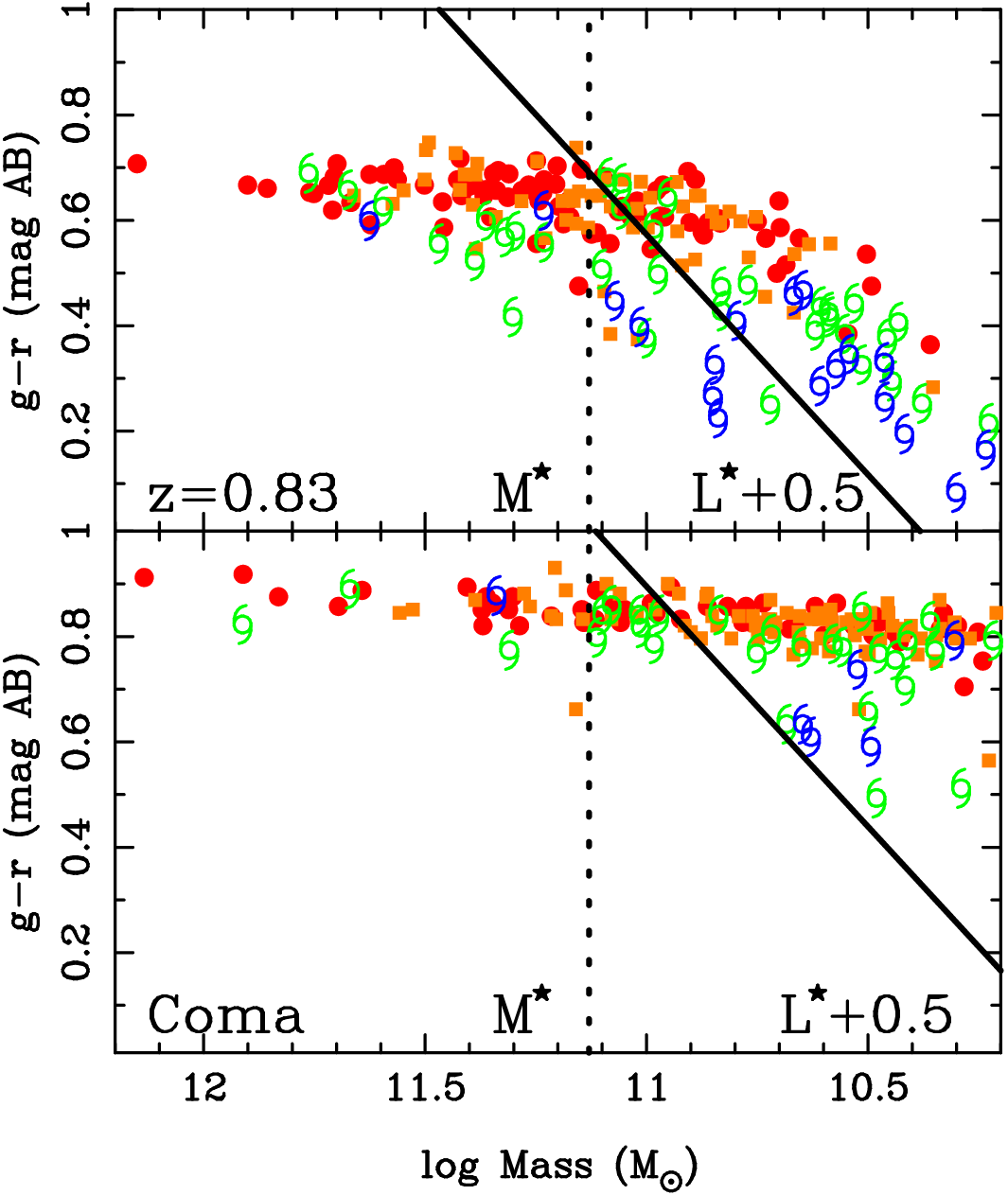}
\end{center}
\caption[color_mass_mag.ps]{Color as a function of mass for galaxies
  at $z=0.83$ (top) and in Coma (bottom).  The mass comes from the
  relation of $M/L_r$ and $g-r$ given by \citet{bell2003}.  The
  symbols represent the same galaxy types as Fig. \ref{fp_masses},
  with the addition of blue spirals for Sc and later types.  The lower
  $M/L$ values of bluer galaxies means that the fraction of early-type
  galaxies at large masses is much higher than the fraction at bright
  luminosities.  The solid line is $L^{\star}+0.5$, a common
  luminosity limit for computing the early-type fraction.  $L^{\star}$
  comes from the Schechter function fits in \citet{beijersbergen2002a}
  and \citet{hansen2005}, and we correct for the luminosity evolution
  of early-type galaxies at $z=0.83$ \cite{holden2005}.  The dotted
  line is $M^{\star}$ from fitting a Schechter function to the mass
  function using the Coma data of \citet{beijersbergen2002a}.  The
  early-type galaxy fraction above $M^{\star}$ is the same in both
  plots.  }
\label{color_ma}
\end{figure}

\section{Mass and the Early-type Fraction}

We find an early-type fraction of 87\% $\pm 6$\% for all galaxies in
Coma with masses above $10^{11.1}\ M_{\sun}$ (errors for the Coma early-type
fractions are computed by using bootstrap re-sampling).  At $z=0.83$,
we find the early-type fraction to be 79\% $\pm 6$\% (82\% $\pm 8$\% for
\ms, 74\% $\pm 8$\% for \cl), a not significant change of 9\%$ \pm 8$\%
from Coma.  In contrast, when using the same data but taking a
luminosity-selected sample of galaxies to $L^{\star} + 0.5$, or $M_r =
-21.1$ mag, we find an early-type fraction of $78 \pm 4$\% for Coma.
At $z=0.83$ we have $62 \pm 6$\% ($65 \pm 8$\% for \ms, $60 \pm 8$\%
for \cl), which is typical for what is found for the cluster samples
in \citet{vandokkum2001} or \citet{holden2004} at the same magnitude
limit.  We note that the $z=0.83$ early-type fractions are corrected
for incompleteness, but this is usually a small, $\le 2$\% effect.

As can be seen in Figure \ref{color_ma}, the mix of galaxy types
selected by a mass cut ($M > M^{\star}$; Fig. \ref{color_ma}, dotted
line) is quite different from that selected by a luminosity cut,
(Fig. \ref{color_ma}, solid line).  To quantify, a blue late-type
galaxy at $L^{\star}$ will have a mass that can be up to 0.4 dex below
$M^{\star}$.  Figure \ref{color_ma} shows that the morphological mix
for a mass selected sample is different from that of a luminosity selected
sample for Coma as well.

The smaller early-type fractions we find at fainter magnitudes means
that, below a certain mass, we should see the early-type fraction
decrease.  We examined the galaxies down to mass equivalent to
$L^{\star} + 0.5$ mag for a red-sequence galaxy, which is $M >
10^{10.9}\ M_{\sun}$.  The early-type fraction for this mass limit is
79\% $\pm 6$\% for Coma and 71\% $\pm 7$\% at $z=0.83$ (78\% $\pm 8$\% \ms\
and 64\% $\pm 8$\% for \cl).  Unfortunately, below this mass threshold
of $M = 10^{10.9}\ M_{\sun}$, the early-type fraction decreases quickly
because we are incomplete, (less than 50\% of early-types at those masses
have redshifts).  The completeness corrections for the early-type
fraction with $M > 10^{10.9}\ M_{\sun}$ are on the order of 5\%.  Thus,
we find no evidence for evolution in the early-type galaxy fraction
even when we examine lower mass samples with two caveats: there is a
larger spread in the measure fractions between \ms\ and \cl, and we
are much more incomplete for redshifts at these masses.

\section{Discussion and Results}

\begin{figure}[thbp]
\begin{center}
\includegraphics[width=3.4in]{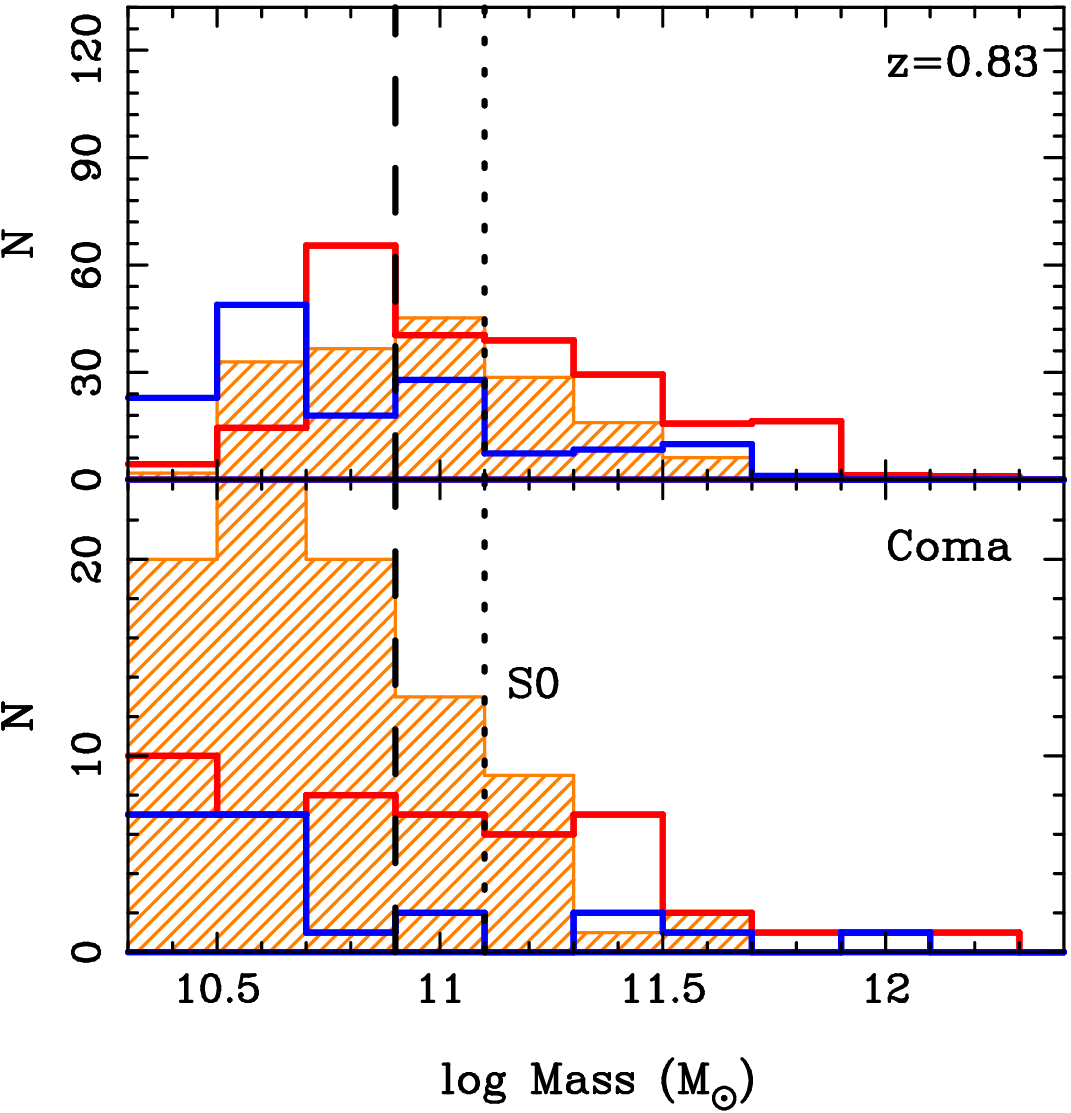}
\end{center}
\caption[mass_dist.ps]{Distribution of masses of the galaxies in
  our two $z=0.83$ clusters (top) and in Coma (bottom).  The red
  histogram shows the elliptical population, the orange hatched
  histogram shows the S0's and S0/a's, and the blue histogram
  represents the late-type galaxies.  The top sample contains all
  $z=0.83$ galaxies with \ia $< 23.5$ mag.  The bottom sample contains
  those Coma members with morphological classifications.  Above
  $M^{\star}$ for Coma (dotted line), the fraction of early-type
  galaxies, E and S0, are the same at both redshifts.  All of the
  evolution in the early-type galaxy fraction observed by others
  occurs at lower masses. The early-type (E/S0) galaxies in the
  $z=0.83$ sample become significantly incomplete ($<$50\%) at masses
  below $10^{10.9}\ M_{\sun}$, marked with a dashed line.  The Coma
  sample is complete to much fainter masses, but we draw the same line
  for comparison.  There is a rapid rise in the number of lower mass
  S0 galaxies in Coma, these masses are coincident with the masses we
  observe for late-type spirals at $z=0.83$.  The bin values for
  $z=0.83$ are weighted by our completeness function, though the
  early-type fractions above $M^{\star}$ are the same without this
  weighting.  }
\label{mass_dist}
\end{figure}

When using magnitude limited samples of galaxies, there is evolution
in the fraction of early-type galaxies over the redshift range 
$z=0$ to $z=1$ \citep{dressler97,lubin98, fasano2000,
  vandokkum2000,vandokkum2001, holden2004, smith2005, postman2005}.
In contrast, Figure \ref{mass_dist} shows that we see minimal
differences in the early-type fraction among galaxies with masses
above $M^{\star}$, a difference of 9\% $\pm 8$\%, in the early-type
galaxy fraction above the mass threshold of $M^{\star}$ between
$z=0.83$ and $z=0$.  The difference in luminosity-selected samples
comes from an increase in the ratio of late-type galaxies, which have
a peak, or modal, mass of $\simeq 10^{10.7}\ M_{\sun}$, and a
corresponding decrease in the number of sub-$M^{\star}$ early-type
galaxies.

A number of recent papers have also claimed to see a lack of evolution
in cluster populations at high masses.  \citet{depropris2003} compared
the evolution in the blue fraction of cluster galaxies in both the
rest-frame $V$ and the observed $K$ band.  The evolution of the blue
fraction is less strong in $K$, a filter that should be more dominated
by massive galaxies than the traditional rest-frame $V$.
\citet{strazzullo2006} also find that the bright end of the
$K$-selected luminosity function does not strongly evolve at redshifts
out to $z\simeq 1.2$.  \citet{tran2004} finds more direct evidence for
the majority of late-type galaxies being low mass from a study of the
recently accreted field galaxies in MS 2053-04 at $z=0.587$.  These
bright, late-type galaxies have star-formation rates similar to that
of field galaxies.  However, based on their mass estimates, these
galaxies will fade onto the lower mass end of the early-type sequence,
presumably to galaxies with less than $10^{11}\ M_{\sun}$.  The excess
of lower mass, late-type galaxies that we observe at high redshift
appear to be at the same mass as the in-falling field galaxy
population observed in \citet{tran2004}. There is a complication in
that the star-formation histories of cluster and field late-type
galaxies may be significantly different \citep{homeier2006}.

\citet{postman2005} found that the fraction of S0 galaxies at almost
all galaxy densities doubles between $z\sim 1$ and $z\sim 0$.  We find
that the fraction of early-type galaxies at high mass, greater than $
M^{\star}$, does not change, within the limits of our errors, over
that same redshift range.  Therefore, most of the evolution observed
in \citet{postman2005} in the galaxy population must occur among lower
mass but still luminous galaxies.  The observed decrease between
$z\sim 1$ and $z\sim 0$ in luminous late-type galaxies could happen in
two ways, the galaxies could be transformed into early-types by a
stopping of star-formation or the galaxies simply fade below the
luminosity limit by $z=0$.  We cannot directly constrain either
scenario with our data.  However, we would like to point out two
observations.  First, \citet{bell2003} finds that $M^{\star}$ for
late-type galaxies is $10^{10.8}-10^{10.9}\ M_{\sun}$ at $z\simeq
0$. If there is no evolution in $M^{\star}$ for late-type galaxies and
a typical-mass late-type galaxy becomes an early-type galaxy through
the truncation of star formation, it should appear as an early-type at
masses below our cutoff of $10^{11.1}\ M_{\sun}$.  At masses of
$10^{10.8}-10^{10.9}\ M_{\sun}$ in Coma, there is a steep rise in the
number of S0 galaxies (see Fig. \ref{mass_dist}).  We note that the
modal mass of late-type galaxies at $z=0.8$ is $10^{10.6}\ M_{\sun}$,
near the mass where the S0 number peaks.  Second, \citet{nelan2005}
predict, based on the ages inferred from absorption line strengths,
that the $10^{10.8}-10^{10.9}\ M_{\sun}$ cluster early-type galaxies
have a median ``age'' of 8--9 Gyr.  This result provides a natural
explanation for the change in the early-type galaxy fraction at these
look-back times.  However, these speculations must be tempered by the
results of \citet{homeier2005} and \citet{crawford2006}, for example.
It is possible that a large fraction of the luminous late-type
galaxies at $z=0.8$ are luminous but low mass galaxies undergoing
starbursts.  These galaxies will simply fade into the dwarf elliptical
population by $z=0$.  Both of these scenarios are directly testable by
measuring the properties of early-type galaxies at lower masses.

Remarkably, most of the massive early-type galaxies appear to be in
place at $z=0.8$, as indicated by the lack of evolution in the
early-type fraction at high masses.  In contrast, at this redshift we
could be seeing the epoch when roughly half of the galaxies with
masses of $(4-8) \times 10^{10}\ M_{\sun}$ have transformed from
late-type galaxies into low-redshift S0 galaxies.  This suggests that
we might have identified the galaxies that will become the bulk of the
$L^{\star}$ early-type population at $z=0$, namely late-type galaxies
with $(4-8) \times 10^{10}\ M_{\sun}$, or $\sim M^{\star}$ for the field
population.

ACS was developed under NASA contract NAS5-32865, and this research was
supported by NASA grant NAG5-7697.  Some of the
data presented herein were obtained at the W.M. Keck Observatory,
which is operated as a scientific partnership among the California
Institute of Technology, the University of California and the National
Aeronautics and Space Administration. The Observatory was made
possible by the generous financial support of the W.M. Keck
Foundation.  The authors wish to recognize and acknowledge the very
significant cultural role and reverence that the summit of Mauna Kea
has always had within the indigenous Hawaiian community.  We are most
fortunate to have the opportunity to conduct observations from this
mountain.

\end{document}